# Synthetic Reflectionless Mode Exceptional Degeneracies via Emergent Local Symmetries

William Tuxbury[1], Lucas Fernandez-Alcazar[1,2,3], Tsampikos Kottos[1]

[1]Wave Transport in Complex Systems Lab, Department of Physics, Wesleyan University, Middletown, CT 06459, USA

[2]Institute for Modeling and Innovative Technology, IMIT (CONICET - UNNE), W3404AAS, Corrientes, Argentina

[3]Physics Department, Natural and Exact Science Faculty, Northeastern University of Argentina, W3404AAS, Corrientes, Argentina

We propose a new kind of physically realizable exceptional point degeneracies (EPDs) corresponding to synthetic reflectionless modes (SRM). These are solutions of an auxiliary wave operator that is defined in synthetic frequency dimensions and describe incoming reflectionless waves onto a Floquet-driven cavity. The SRM-EPD emerges as a consequence of a spontaneous *local* $\mathcal{PT}$-symmetry imposed on the auxiliary operator via appropriate Floquet driving. Its presence signifies the possibility to design wavefronts and time-modulated schemes with up/down targeted frequency conversion and flat transmission spectra. The theory is validated via simulations with driven RF resonators.

*Introduction* - The physics of low (geometrical) dimensional systems is often constrictive, eliminating the possibility to observe a variety of exotic phenomena including Anderson Metal Insulator transition [1][2][3][4], spontaneous symmetry-breaking phase transition in the ferromagnetic Ising model [5][6][7][8], nontrivial topological effects [9][10][11][12] and more. To bypass this hardship, the research community has recently developed the concept of synthetic dimensions, which allows extending investigations beyond the apparent geometrical dimensionalities of a physical system [13][14][15][16]. Synthetic dimensions have been formed by coupling states labeled by different internal degrees of freedom to form a lattice. These degrees of freedom may be the frequency, spin, linear momentum, orbital angular momentum, etc. Such synthetic space constructions also enable new possibilities for manipulating these internal degrees of freedom, which are of significant potential importance for applications such as communications and information processing.

Along these lines, synthetic frequency lattices, formed by engaging an appropriately designed periodic (Floquet) driving of cavity modes, can be used for implementing multi-dimensional convolution operations [17], investigating higher-dimensional topological physics in low-dimensional physical structures [13][14][15][16][18], or as a means to enforce non-reciprocal transport and up/down frequency conversion [19][20][21][22]. In many of these cases, it is desirable to develop wavefront-shaping schemes that eliminate reflected waves which might have serious damaging effects or a targeted up/down frequency conversion with a flattened transmission profile, etc.

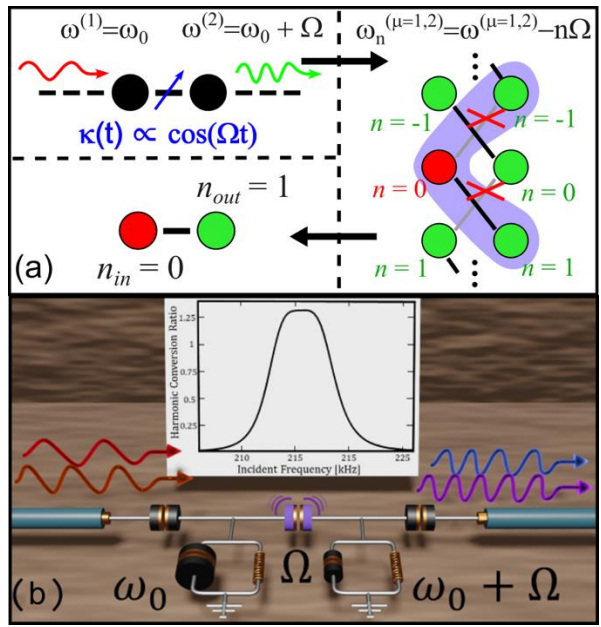

FIG 1. (a) Coupled Mode Theory schematic. (upper left) The dimer with modulated coupling, corresponding frequency detuning, and TLs coupled to either site. (right) The equivalent synthetic frequency lattice whose eigenstates are the reflection-zeros. The red site depicts effective gain due to coupling the incoming wave, while green sites depict effective loss due to coupling the outgoing wave. The red cross out indicates those replicas are decoupled due to frequency mismatch. (lower left) The resulting auxiliary system after approximation. The $\mathcal{PT}$-symmetric dimer connects the input wave to the first harmonic, enabling an SRM-EPD and resulting in a quartic frequency conversion. (b) Illustration of the physical system of two coupled LC resonators with periodically modulated capacitive coupling and capacitive coupling with two transmission lines. The frequency detuning between the two resonators is equal to the modulation frequency and results in a frequency conversion exhibiting a quartic lineshape.

Structured wavefront protocols that offer unprecedented control of waves in complex settings have been developing during the last decade for a variety of applications including imaging, wireless communications, etc., [23][24][25]. These works, however, were solely focused on wave scattering control from static systems. Thus, wavefront-shaping schemes in the case of synthetic frequency lattices—or any other type of synthetic dimensions for that matter— were largely left unexplored.

In fact, some of the existing wavefront-shaping schemes have been used for the implementation and study of exotic non-Hermitian singularities known as exceptional point degeneracies (EPDs) [26][27][28][29]. These are singularities in the parameter space of the wave system at which two or more discrete eigenvalues and the associated eigenstates of a non-Hermitian wave operator coalesce. Although EPDs have initially been explored in the framework of resonant modes [30][31][32][33] [34][35][36][37][38][39][40][41][42][43], their investigation recently has been extended to non-Hermitian auxiliary wave operators describing scattering processes with designated incoming channels connected to the scattering system modeled as gain and outgoing channels (or system losses) modeled as losses [44][45][46][47][48][49][50][51][53]. Outcomes of these endeavors include EPDs of linear and nonlinear coherent perfect absorption (CPA) modes [46][47][53][54], or its generalization associated with reflectionless scattering states and flattened transmission spectra [48][49][50][51][52]. Another recent exploration of cavities at CPA-EPD parameter configuration revealed their use as robust 50-50 power splitters for any input signal (apart from zero-measure wavefronts corresponding to a CPA wavefront) [55].

Here, we foster the elevation of these wavefront-shaping techniques and the implementation of EPDs to synthetic frequency spaces associated with Floquet-driven systems. We showcase this new family of EPDs in the spectrum of an auxiliary wave operator that describes incoming reflectionless waves from the fundamental frequency onto a two-mode cavity whose modal coupling is periodically modulated with a driving frequency that is equal to the frequency detuning between the two modes, Fig. 1a. The resulting synthetic reflectionless mode (SRM) EPD emerges as a consequence of a spontaneous hidden parity-time ($\mathcal{PT}$) symmetry imposed on the auxiliary operator via appropriate Floquet driving. We show that the injected wave is routed to a targeted harmonic while the outgoing transmission spectrum acquires a flattened band profile due to the aforementioned EPD. The theory is confirmed via simulations with RF Floquet resonators, Fig. 1b. The proposed spectrum management might find various applications from targeted frequency conversion to indoor wireless communications (frequency-shifting of signals to specific users via modulated RIS).

*Synthetic Reflectionless Modes* - We employ coupled mode theory (CMT) as a generic framework to demonstrate the implementation of Floquet wavefront schemes in synthetic frequency lattices. We will be focusing on a scheme that highlights SRM and the formation of SRM-EPDs via hidden symmetries due to the importance of suppressing backscattering occurrences. We will also leverage the methodology to achieve broadened frequency targeted conversion. As a prototype example, we introduce the following scattering system consisting of two resonators with a periodically modulated coupling between them and with each resonator coupled to its corresponding transmission line (Fig. 1a). The natural frequencies of the resonators are mismatched by a detuning $\Delta$. The time-domain equations of motion for this system are

$$i\frac{d}{dt}|\Psi\rangle = H_{eff}|\Psi\rangle + iD|S^+\rangle \quad (1a)$$
$$|S^-\rangle = -|S^+\rangle + D^T|\Psi\rangle \quad (1b)$$

where $|\Psi\rangle$ and $|S^\pm\rangle$ are time-dependent $2\times 1$ vectors, whose components represent the amplitudes of the resonator modes and incoming $(+)$ or outgoing $(-)$ waves in the transmission lines (TL), respectively. The connection between the wave in the TLs and within the system is encoded through the coupling matrix $D$. For simplicity, by assuming symmetric, non-dispersive coupling to the TLs $D = \sqrt{2\gamma_e}\mathbb{I}_2$, where $\mathbb{I}_2$ is the $2 \times 2$ identity matrix and $\gamma_e$ is the coupling to the TLs. Equation (1a) describes the internal dynamics of the open system, subject to a source, by its effective Hamiltonian $H_{eff} \equiv H_0(t) - \frac{i}{2}DD^T$. Here, the time-dependent Hamiltonian of the isolated system is $H_0(t) = \begin{pmatrix} \omega^{(1)} & \kappa(t) \\ \kappa(t) & \omega^{(2)} \end{pmatrix}$, where $\omega^{(1)} = \omega_0$ and $\omega^{(2)} = \omega_0 + \Delta$, and the non-Hermitian term captures the system coupling to the TLs as a surrogate loss term. The resonator coupling in the off diagonal is time periodic $\kappa(t) = \kappa(t + 2\pi/\Omega)$ and will take the explicit form $\kappa(t) = \kappa_0 + 2\kappa_1 \cos\Omega t$, where $\kappa_0$ and $\kappa_1$ are parameters. Meanwhile, the diagonal components encode the detuned frequencies of the isolated resonators. A schematic of this system is shown in the upper-left panel of Fig. 1a. Equation (1b) is an input/output relationship that encodes flux conservation between the TLs and the system.

Since Eq. (1a) describes a time-periodic linear system, it cannot be directly transformed into the frequency domain. Rather, the time-dependent system is first mapped onto an equivalent time-independent, albeit infinite-dimensional, synthetic lattice. To this end, we expand the effective Hamiltonian and wave variables in Fourier series $H_{eff} = \sum_l H_F^{(l)} e^{-il\Omega t}$ and $|F\rangle = \sum_n e^{-i(\omega + n\Omega)t}|f_n\rangle$ centered at the incident reference frequency $\omega$, respectively. Here, $F$ stands for either $\Psi$ or $S^\pm$, while $f_n$ can be either $\psi_n$ or $s_n^\pm$. Thus, $|f_n\rangle$ is a vector of the Fourier amplitudes of the $n^{th}$ harmonic of $|F\rangle$. This enables us to recast Eq. (1) in the extended space of harmonics,

$$\omega\vec{\psi} = H_F\vec{\psi} + i\widetilde{D}\vec{s}^+ \quad (2a)$$
$$\vec{s}^- = -\vec{s}^+ + \widetilde{D}^T\vec{\psi} \quad (2b)$$

where the vector $\vec{f} \equiv (\ldots \ |f_1\rangle \ |f_0\rangle \ |f_{-1}\rangle \ \ldots)^T$ contains the wave amplitudes over all harmonics and $\widetilde{D} = \text{diag}(\ldots, D, D, \ldots)$ is the block-diagonal matrix with $D$ along the diagonal, which describes coupling to each harmonic of the TLs (see SM A for details). The structure of the synthetic lattice described by $H_F$, also called the Floquet ladder, can be understood as a sequence of replicas of the original system's static elements (including TLs), each incremented in frequency by the modulation rate and interconnected by coupling elements determined by the Fourier coefficients of the modulation. For example, coupling $\kappa(t) = \kappa_0 + 2\kappa_1 \cos\Omega t$ between the first and second resonators of the physical system induces coupling $\kappa_1$ between the first and second resonators of adjacent replicas. Each replica corresponds to a different harmonic, e.g., a wave incident at frequency $\omega$, exiting the synthetic lattice through the next replica in the ladder, will then transmit from the physical system with frequency $\omega + \Omega$. Frequency conversion between harmonics can be numerically computed via the Floquet scattering matrix after truncation of the synthetic lattice (seven total replicas were included in all simulations),

$$S_F = -\widetilde{\mathbb{I}}_M + i\widetilde{D}^T G_F \widetilde{D} \quad (3)$$

where $G_F = [\omega\widetilde{\mathbb{I}}_N - H_F]^{-1}$ is the Green's function of the synthetic lattice with $\widetilde{\mathbb{I}}_N$ and $\widetilde{\mathbb{I}}_M$ being the identity matrices with corresponding dimensionality of $H_F$ and $S_F$, respectively.

To devise a scheme that will enable SRM and targeted frequency conversion with a broadened lineshape, we impose as a requirement reflectionless scattering boundary conditions at the corresponding incident harmonic and TL. To this end, we introduce an auxiliary synthetic lattice, depicted in the right panel of Fig. 1a, where the effective loss (indicated by green sites) due to its TL coupling is replaced with an effective gain (indicated by the red site) at the input. By enforcing reflectionless scattering boundary conditions, the wave operator associated with this auxiliary synthetic lattice $H_{SRZ}$ can be used to collapse Eqs. (2) to an equivalent eigenvalue problem,

$$\omega\vec{\psi} = H_{SRZ}\vec{\psi} \quad (4)$$

Therefore, synthetic reflectionless solutions are encoded as eigenstates of $H_{SRZ}$ for a wave injected at the associated eigenfrequency, i.e., reality of the eigenfrequencies is a criterion for SRMs (see SM C for details). While an SRM guarantees zero reflection at the input TL and frequency harmonic, the associated scattering boundary conditions do not define the channel(s) to which the wave is transmitted.

It turns out the determinant of the Floquet reflection matrix $\hat{r}_F$, given by the associated subblock of $S_F$, is related to the determinants of the dynamical and auxiliary wave operators of the synthetic lattice, $M_F$ and $M_{SRZ}$ respectively,

$$\det \hat{r}_F = \pm \frac{\det M_{SRZ}}{\det M_F} \quad (5)$$

where the sign is determined by the choice of representation for $M_{SRZ}$ and dimensionality of $\hat{r}_F$ (see SM C for details). While Eq. (5) is generic, in the case of CMT modeling $M_F \equiv \omega \tilde{\mathbb{I}}_N - H_F$ from Eq. (2) and $M_{SRZ} \equiv \omega \tilde{\mathbb{I}}_N - H_{SRZ}$ from Eq. (4). For a single input channel, the Floquet reflection matrix simplifies to a scalar, $\hat{r}_F \equiv r_F$. Furthermore, if $M_{SRZ}$ possesses a second-order zero on the real axis at $\omega = \omega_*$ – corresponding to a real eigenvalue degeneracy in the spectrum of $H_{SRZ}$ – that is also isolated from resonance poles of the system (zeros of $M_F$), then the Floquet reflection coefficient will exhibit anomalous scaling behavior in proximity to the degenerate zero, $r_F \sim (\omega - \omega_*)^2$. This leads to a quartic lineshape of the reflectance,

$$R \equiv |r_F|^2 \sim \nu^4 \quad (6)$$

where $\nu \equiv \omega - \omega_*$ is frequency detuning from the degeneracy. We have found that this degeneracy in the spectrum of $H_{SRZ}$ for the driven dimer occurs at $\omega_* = \omega_0$ when $\kappa_1 = \gamma_e$, and $\Delta = \Omega \gg \kappa_1$. The previous inequality guarantees an optimal frequency up-conversion from $\omega \to \omega + \Omega$ for incident waves with frequencies $\omega \approx \omega_0$. In the case of minimal loss to parasitic channels, Eq. (6) predicts a subsequent broadening in the transmittance to the adjacent harmonic. A similar result applies for down-conversion in the case that $\Delta = -\Omega$.

The following results are presented with $\omega_0 = 0$ as reference, $\Omega = 1$ as scale and $\Delta = \Omega$ to achieve frequency up-conversion. In Fig. 2a, the solid curves depict the real and imaginary parts of the reflection zero eigenfrequencies for $\kappa_0 = 0$ as a function of the modulation strength $\kappa_1/\gamma_e$, clearly showing the bifurcation characteristic of an EPD. In Fig. 2b, the blue and red curves illustrate reflection back to the input harmonic and transmission to the $n = 1$ harmonic, respectively; from the near-total frequency conversion, it is evident that transmission to parasitic channels is negligible. The inset shows the predicted quartic scaling of the reflection. Subtle deviation from an exact degeneracy at the bifurcation in Fig. 2a, along with the finite (albeit insignificant $\sim 10^{-5}$) reflection plateau in Fig. 2b, are consequences of the finite ratio of $\gamma_e/\Omega = 5 \times 10^{-3}$. At the same subfigures, we also report results (dashed curves) corresponding to $\kappa_0 = 3\gamma_e$, ensuring that $\kappa(t)$ is strictly positive for all displayed values of $\kappa_1$. While the bifurcation is degraded in Fig. 2a, still one can see strong evidence of an EPD in the parametric analysis of the Floquet modes. At the same time, the transmission and reflection characteristics are qualitatively unaffected, see Fig. 2b. The inset shows an increase in the reflection plateau, though it is still very small ($\sim 10^{-3}$).

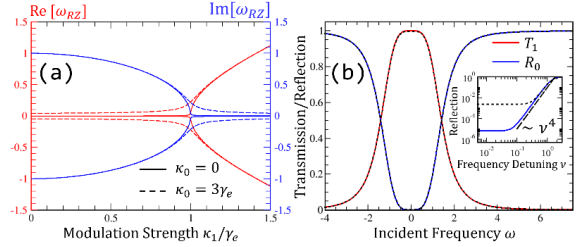

FIG 2. (a) The red (blue) lines represent the real (imaginary) parts of the SRM eigenfrequencies for two values of static coupling $\kappa_0$. (b) The red (blue) line shows transmission (reflection) to the first harmonic (incident harmonic). The inset depicts quartic reflectance scaling. In both subfigures, the solid (dashed) lines correspond to scenarios of modulated coupling with $\kappa_0 = 0$ ($\kappa_0 = 3\gamma_e$). For $\kappa_0 = 3\gamma_e$, although the bifurcation cusp in (a) is notably degraded, the results in (b) remain qualitatively unaffected and are therefore shown in black for visual clarity.

The conditions (system parameters) under which SRM-EPDs are formed call for a physical explanation. Frequency up- (down-)conversion to the second TL can be facilitated by tuning $\Delta = \Omega$ ($\Delta = -\Omega$) to match the resonance frequency of the output site of the synthetic lattice, i.e., replica $n_{out} = 1$ ($n_{out} = -1$). For sufficiently large modulation frequency ($\Omega \gg \kappa_0, \kappa_1$), the frequency mismatch suppresses transmission to parasitic channels, as indicated by the red strikeouts in the right panel of Fig. 1a. Focusing only on contributing sites, what remains of the auxiliary synthetic lattice is the $\mathcal{PT}$-symmetric dimer with coupling $\kappa_1$ and gain/loss $\gamma_e$, as depicted in the lower-left panel of Fig. 1a. Such a physical picture unveils the existence of a *local* $\mathcal{PT}$-symmetry [56][57][58] present in the synthetic frequency space of our Floquet two-mode cavity and is responsible for the formation of the SRM-EPD. Specifically, by setting $\kappa_1 = \gamma_e$, the $\mathcal{PT}$-symmetric dimer is brought to a symmetry-breaking transition, inducing an exceptional point degeneracy (EPD) of SRMs and a consequent broadening of the reflectance and transmission spectrum. A detailed analysis, involving a decimation approach [59] that maps the original infinite dimensional synthetic frequency lattice to an effective $\mathcal{PT}$-symmetric dimer system, can be found at the SM A.

*Implementation of SRM to an RF Cavity* - Next, we implement the above methodology and system architecture to a physical RF cavity, see Fig. 1b. The corresponding circuit diagram is shown in Fig. 3a and consists of two parallel LC tanks with inductance $L$ and capacitances $C_1$ and $C_2$ respectively. Each LC tank is capacitively coupled to transmission lines (TLs) with characteristic impedance $Z_0 = 50\ [\Omega]$ and capacitances $C_1^e$ and $C_2^e$ respectively. The LC tanks are then coupled to one another by a time-modulated capacitance $C^\kappa(t) = C_0^\kappa + 2C_1^\kappa \cos \Omega t$, with $0 \leq \eta \equiv C_1^\kappa/C_0^\kappa < 1/2$ to keep the capacitance strictly positive for physical consistency (for a physical implementation see [60]).

The circuit dynamics are determined by Kirchhoff's voltage and current conservation laws in the time domain. To analyze scattering in the time-periodic circuit, following a similar approach as with the CMT, we have derived a frequency domain formulation of the problem (see SM B for details),

$$M_F \vec{\Phi} = A_F \vec{V}^+ \quad (7a)$$
$$\vec{V}^- = -\vec{V}^+ + \widetilde{W}\vec{\Phi} \quad (7b)$$

In the first equation, $\vec{\Phi}$ is a vector of voltages at each harmonic in the TLs and LC tanks, and $\vec{V}^+$ is a vector containing voltage amplitudes of the incident waves in the TLs at each excited harmonic, e.g., $\ldots, \omega - \Omega, \omega, \omega + \Omega, \ldots$. The wave operator $M_F$ is a block tridiagonal matrix that encodes intra-harmonic and inter-harmonic interactions in its diagonal and off-diagonal blocks, respectively. Meanwhile, $A_F$ accounts for coupling input waves from the TLs and $\widetilde{W}$ is a matrix such that $\vec{V} \equiv \widetilde{W}\vec{\Phi}$, where $\vec{V}$ is a vector that represents the voltages at each harmonic in the TLs. Therefore, the second equation simply expresses how voltages in the TLs are decomposed into incoming and outgoing components, where the latter are represented by $\vec{V}^-$.

For the analysis of SRMs, Eq. (7a) can be manipulated to result in an auxiliary wave operator whose null vectors are the reflection zeros (see SM C for details),

$$M_{SRZ}\vec{\Phi} = A_{SRZ}\vec{V}_{SRZ} \quad (8)$$

where $\vec{V}_{SRZ}$ is identical to $\vec{V}^+$, aside from the one component, which is the incident voltage amplitude at the input TL and harmonic and is replaced by the corresponding scattered voltage amplitude. Hence, the Floquet reflectionless scattering boundary conditions are met when $\vec{V}_{SRZ} = 0$, corresponding to zero reflection at the input channel(s) and no incident wave at all other channels. Consequently, non-trivial solutions for $\Phi$ exist at singular points of $M_{SRZ}$, so the circuit will support SRMs when $\det M_{SRZ}(\omega) = 0$ at real frequency $\omega \in \mathbb{R}$. It can be demonstrated that the matrix representations in Eq. (8) may be interpreted physically in relation to their counterparts in Eq. (7a). Specifically, changing the sign of the characteristic impedance at the input TL and harmonic will turn $M_F$ and $A_F$ into $M_{SRZ}$ and $A_{SRZ}$, respectively.

Guided by the symmetry principles prescribed by the CMT modeling, the circuit parameters were iteratively refined to induce a degeneracy of zeros in Eq. (8) near to the real axis (see SM D for details). In the simulations, which follow the standard positive frequency convention used in circuit analysis, we verified the stability of the system by confirming that its resonance poles (i.e., zeros of Eq. (7a)) lie above the real axis. The results for two configurations of the circuit parameters are presented in Figs. 3b-d. In Fig. 3b, the real and imaginary parts of the complex reflection zeros are plotted as a function of the modulation strength $\eta$, displaying a distinct bifurcation where the zeros coincide. The solid and dashed curves correspond to the parameter configurations used to obtain the results presented in Figs. 3c and 3d, respectively. In Figs. 3c and 3d, the main panels show transmission to the $n = 1$ harmonic of the second TL in red, reflection back to the $n = 0$ harmonic of the first TL in blue, and net transmission to all other channels in orange. Since the reflection zeros are not strictly real, the insets show that quartic scaling is preserved in the differential reflectance as a function of the incident frequency detuning. The results of Fig. 3c indicate an amplified transmission to the target channel due to the modulation of the capacitor. Meanwhile, the results of Fig. 3d use a modulation frequency less than one third of that used in Fig. 3c, which reduced the target transmission to below one but also enhanced transmission to parasitic channels and distorted the quartic lineshape in the transmission.

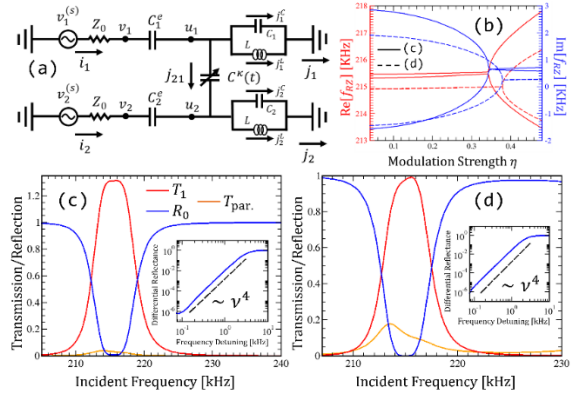

FIG 3. (a) Circuit diagram. (b) Real (red lines) and imaginary (blue lines) parts of synthetic reflection-zero frequencies as a function of the modulation strength. Solid and dashed curves correspond to parameter configurations of (c) and (d), respectively. (c) (d) Red (blue) curves are transmission (reflection) to the first excited harmonic (to the incident harmonic), denoted by $T_1$ ($R_0$). Orange curves show parasitic transmission $T_{\text{par.}}$ to all other channels. The insets display quartic scaling of the differential reflectance. In (c), $T_1 > 1$ due to parametric amplification, while in (d), the driving frequency is reduced, resulting $T_1 < 1$ and elevated $T_{par.}$

*Conclusions* – We have introduced wavefront shaping in synthetic frequency dimensions of periodically modulated cavities and identified a new kind of physically realizable exceptional point degeneracies corresponding to synthetic reflectionless modes. These SRM-EPDs are a consequence of a tailored Floquet driving that enforces a hidden $\mathcal{PT}$ symmetry to an auxiliary operator that describes scattering states with zero reflection. We have shown that their implementation results in targeted up/down frequency conversion with flat transmission profiles. Due to a lack of space, we are unable to discuss some more results and possible experimental demonstrations. The wealth of applications there is to be gained from the implementation of wavefront shaping and EPDs in synthetic frequency dimensions can be extended beyond SRMs or even to synthetic lattices associated with the manipulation of other internal degrees of freedom.

*Acknowledgments* – We acknowledge partial support from MPS Simons Collaboration via grant No. SFI-MPS-EWP-00008530-08. W.T. and T.K. also acknowledge partial support from DOE DE-SC0024223, MURI ONR-N000142412548, NSF-RINGS ECCS 2148318 and from the BSF2022158.

Synthetic Reflectionless Mode Exceptional Degeneracies via Emergent Local Symmetries
William Tuxbury[1], Lucas Fernandez-Alcazar[1,2,3], Tsampikos Kottos[1]
[1]Wave Transport in Complex Systems Lab, Department of Physics, Wesleyan University, Middletown, CT 06459, USA
[2]Institute for Modeling and Innovative Technology, IMIT (CONICET - UNNE), W3404AAS, Corrientes, Argentina
[3]Physics Department, Natural and Exact Science Faculty, Northeastern University of Argentina, W3404AAS, Corrientes, Argentina

## Supplementary Material

### Supplementary Material A: Synthetic Lattice

In this section of the supplementary material, we provide a concise derivation of the mapping from the time-periodic system (1) to an equivalent time-independent synthetic lattice. Using this mapping, we shall write the resulting synthetic scattering matrix (subsection A.1). Finally (subsection A2) we will show, using a decimation method, the local parity-time ($\mathcal{PT}$) symmetry that characterizes the driven dimer.

The synthetic lattice, or Floquet ladder, consists of a sequence of replicas that resemble the original system, whose natural frequencies of each replica are incremented by the modulation frequency corresponding to different harmonics, and whose coupling structure is determined by the Fourier components of the modulation. Assuming a periodic modulation of the system $H_{eff}(t) = H_{eff}\left(t + \frac{2\pi}{\Omega}\right)$, we can express the effective Hamiltonian using a Fourier series $H_{eff}(t) = \sum_l H_F^{(l)} e^{-il\Omega t}$ (sum over all integers), where the Fourier coefficients are given by $H_F^{(l)} = \frac{\Omega}{2\pi}\int_0^{2\pi/\Omega} dt\, e^{il\Omega t} H_{eff}(t)$. Together with a monochromatic incident wave $|S^+\rangle = e^{-i\omega t}|s_0^+\rangle$, the system response is $|\Psi\rangle = \sum_n e^{-i(\omega+n\Omega)t}|\psi_n\rangle$. Inserting these expansions into (1a), canceling the common factor $e^{-i\omega t}$, and applying $\frac{\Omega}{2\pi}\int_0^{2\pi/\Omega} dt\, e^{in\Omega t}$, we obtain,

$$(\omega + n\Omega)|\psi_n\rangle = \sum_l H_F^{(l)}|\psi_{n-l}\rangle + \delta_{n0} iD|s_0^+\rangle \quad (A1)$$

which follows from the orthogonality relation $\int_0^{2\pi/\Omega} e^{i(n-m)\Omega t} = \frac{2\pi}{\Omega}\delta_{nm}$ and is aided by the assumption that $D$ is non-dispersive. Since this holds for each harmonic $n$, we introduce an extended-space notation to express a matrix formulation,

$$\omega\vec{\psi} = H_F\vec{\psi} + i\widetilde{D}\vec{s}^+ \quad (A2)$$

where $\vec{f} \doteq (\ldots\ |f_1\rangle\ |f_0\rangle\ |f_{-1}\rangle\ \ldots)^T$ for $f$ either $\psi$ or $s^{\pm}$, thus generalizing to incident waves at arbitrary harmonics such that $|S^{\pm}\rangle = \sum_n e^{-in\Omega t}|s_n^{\pm}\rangle$. Furthermore, $\tilde{A}$ denotes a block-diagonal matrix with repeated blocks $A$, and $H_F$ is the effective Hamiltonian of the synthetic lattice with blocks,

$$[H_F]_{pq} = H_F^{(q-p)} - \delta_{pq}\mathbb{I}_N p\Omega \quad (A3)$$

Here, $\mathbb{I}_N$ is an identity matrix of the same dimensionality as $|\Psi\rangle$. Notice, the second term encodes the frequency shift between replicas.

### 1. Floquet Scattering Matrix

Together with the analog of (1b) in the extended-space representation,

$$\vec{s}^- = -\vec{s}^+ + \widetilde{D}^T\vec{\psi} \quad (A4)$$

we can write the Green's function and scattering matrix in this Floquet ladder of harmonics,

$$G_F \equiv [\omega\widetilde{\mathbb{I}}_N - H_F]^{-1} \quad (A5)$$
$$S_F \equiv -\widetilde{\mathbb{I}}_M + i\widetilde{D}^T G_F \widetilde{D} \quad (A6)$$

where $\mathbb{I}_M$ is an identity matrix of the same dimensionality as $|S^{\pm}\rangle$. The frequency-domain response is given by $\vec{\psi} = iG_F\widetilde{D}\vec{s}^+$ and $\vec{s}^- = S_F\vec{s}^+$.

When proceeding numerically, $H_F$ must be truncated before $G_F$ can be computed. If only a few harmonics are involved in the modulation, a small number of replicas will typically suffice. Fewer replicas will be required when either the modulation strength is weak (which reduces inter-harmonic coupling), or the modulation frequency is large (which increases frequency mismatch between replicas). For both CMT and circuit modeling (discussed in SM B), a total of seven replicas were used in all simulations.

## 2. Decimation-Based Reduction of the Floquet Lattice to an Effective $\mathcal{PT}$-symmetric Dimer under SRM Conditions

In this section, we provide a detailed analysis of the mapping between the infinite-dimensional Floquet lattice and an effective $\mathcal{PT}$-symmetric dimer. To this end, we use the decimation technique to reduce the dimensionality of the Floquet lattice to an effective dimer that describes the system's response. By analyzing the synthetic reflectionless modes, we obtain an auxiliary wave operator for the dimer that respects $\mathcal{PT}$-symmetry.

Let us start by discussing the case of a periodically time-modulated Hamiltonian, $H(t) = H(t + \frac{2\pi}{\Omega})$, such that

$$H(t) = \begin{pmatrix} \omega^{(1)} & \kappa(t) \\ \kappa(t) & \omega^{(2)} \end{pmatrix}, \qquad (A7)$$

where $\omega^{(2)} = \omega^{(1)} + \Omega$ and only the coupling between resonators varies with time as $\kappa(t) = \kappa_0 + 2\kappa_1 \cos(\Omega t)$. Then, the Fourier series for the effective Hamiltonian resulting after including the corrections due to leads, is given by $\widetilde{H}(t) \equiv H(t) - i\frac{DD^T}{2} = \sum_l e^{-il\Omega t} \cdot H_F^{(l)}$, with (matrix) coefficients

$$H_F^{(0)} = \begin{pmatrix} \omega_r & \kappa_0 \\ \kappa_0 & \omega_r + \Omega \end{pmatrix}; \quad H_F^{(\pm 1)} = \begin{pmatrix} 0 & \kappa_1 \\ \kappa_1 & 0 \end{pmatrix}; \quad H_F^{(l)} = 0 \text{ for } |l| \geq 2. \qquad (A8)$$

Here $\omega_r \equiv \omega^{(1)} - i\gamma_e$ and $D = \sqrt{2\gamma_e}\mathbb{I}_2$. Then, the Floquet Hamiltonian in Eq. (A2) has a block tridiagonal structure

$$\omega \begin{pmatrix} \vdots \\ |\psi(\omega_1)\rangle \\ |\psi(\omega_0)\rangle \\ |\psi(\omega_{-1})\rangle \\ \vdots \end{pmatrix} = \begin{pmatrix} \ddots & & & 0 \\ & H_F^{(0)} - \Omega I & H_F^{(1)} & 0 \\ & H_F^{(-1)} & H_F^{(0)} + 0 \cdot I & H_F^{(1)} \\ & 0 & H_F^{(-1)} & H_F^{(0)} + \Omega I \\ 0 & & & \ddots \end{pmatrix} \begin{pmatrix} \vdots \\ |\psi(\omega_1)\rangle \\ |\psi(\omega_0)\rangle \\ |\psi(\omega_{-1})\rangle \\ \vdots \end{pmatrix} + i\widetilde{D} \begin{pmatrix} \vdots \\ 0 \\ |S^+(\omega_0)\rangle \\ 0 \\ \vdots \end{pmatrix}; \qquad (A9)$$

where $\omega_n \equiv \omega + n\Omega$. To illustrate the role of the modulation frequency in the frequency mismatch between Floquet replicas, we proceed by explicitly writing the equation above

$$\omega \begin{pmatrix} \vdots \\ \psi_{n=1,1} \\ \psi_{n=1,2} \\ \psi_{n=0,1} \\ \psi_{n=0,2} \\ \psi_{n=-1,1} \\ \psi_{n=-1,2} \\ \vdots \end{pmatrix} = \begin{pmatrix} \omega_r - \Omega & \kappa_0 & & \kappa_1 & & & \\ \kappa_0 & \omega_r & \kappa_1 & & & & \\ & \kappa_1 & \omega_r & \kappa_0 & & \kappa_1 & \\ \kappa_1 & & \kappa_0 & \omega_r + \Omega & \kappa_1 & & \\ & & & \kappa_1 & \omega_r + \Omega & \kappa_0 & \\ & & \kappa_1 & & \kappa_0 & \omega_r + 2\Omega \end{pmatrix} \begin{pmatrix} \vdots \\ \psi_{n=1,1} \\ \psi_{n=1,2} \\ \psi_{n=0,1} \\ \psi_{n=0,2} \\ \psi_{n=-1,1} \\ \psi_{n=-1,2} \\ \vdots \end{pmatrix} + i\widetilde{D} \begin{pmatrix} \vdots \\ 0 \\ 0 \\ S^+_{n=0,1} \\ 0 \\ 0 \\ 0 \\ \vdots \end{pmatrix}, \qquad (A10)$$

where the subindex $(n,j)$ represents the Floquet replica $n$ and site number $j = 1,2$. This representation facilitates the identification of impedance-matched blocks within the Floquet Hamiltonian. Notice that such blocks involve wave amplitudes at different Floquet replicas, specifically, the diagonal blocks couple field amplitudes $\psi_{n,2}$ with $\psi_{n-1,1}$.

While there exist infinitely many such matched pairs throughout the lattice, only the amplitude $\psi_{n=0,1}$, which is directly excited by the incident wave, is impedance matched with $\psi_{n=1,2}$.

We next apply a decimation technique, commonly used in condensed matter frameworks [59], to systematically eliminate from the description all modes that neither satisfy the impedance matching conditions nor are directly connected to the source. Specifically, we begin by decimating the non-central Floquet blocks with $|n| \geq 2$, which are not directly influenced by the incident wave. This procedure yields corresponding self-energies, $\Sigma_{-2}$ and $\Sigma_{+2}$, which effectively renormalize the blocks associated with the

$n = -1$ and $n = +1$ replicas, respectively. These self-energies are assumed to exist and be computable. Then,

$$\omega \begin{pmatrix} |\psi(\omega_1)\rangle \\ |\psi(\omega_0)\rangle \\ |\psi(\omega_{-1})\rangle \end{pmatrix} = \begin{pmatrix} H_F^{(0)} - \Omega I + \Sigma_{+2} & H_F^{(1)} & 0 \\ H_F^{(-1)} & H_F^{(0)} & H_F^{(1)} \\ 0 & H_F^{(-1)} & H_F^{(0)} + \Omega I + \Sigma_{-2} \end{pmatrix} \begin{pmatrix} |\psi(\omega_1)\rangle \\ |\psi(\omega_0)\rangle \\ |\psi(\omega_{-1})\rangle \end{pmatrix}$$

$$+ i \begin{pmatrix} 0 \\ D|S^+(\omega_0)\rangle \\ 0 \end{pmatrix}, \qquad (A11).$$

Next, we decimate the block $n = -1$. Then, the directly connected block $n = 0$, will be corrected as

$$\widetilde{H}_F^{(0)} \equiv H_F^{(0)} + \Sigma_{-1}; \quad \Sigma_{-1} \equiv H_F^{(1)} \cdot \left(\omega I - \left(H_F^{(0)} + \Omega I + \Sigma_{-1}\right)\right)^{-1} \cdot H_F^{(-1)}, \quad (A12)$$

where the inverse matrix is given by

$$(\cdot)^{-1} = \frac{1}{\det(\cdot)} \begin{pmatrix} \omega - [\omega_r + 3\Omega + (\Sigma_-)_{2,2}] & \kappa_0 + (\Sigma_-)_{1,2} \\ \kappa_0 + (\Sigma_-)_{2,1} & \omega - [\omega_r + 2\Omega + (\Sigma_-)_{1,1}] \end{pmatrix}$$

$$\det(\cdot) = [\omega - [\omega_r + 3\Omega + (\Sigma_-)_{2,2}]] \cdot [\omega - [\omega_r + 2\Omega + (\Sigma_-)_{1,1}]] - (\kappa_0 + (\Sigma_-)_{2,1}) \cdot (\kappa_0 + (\Sigma_-)_{1,2})$$

It is important to note that a large modulation frequency $\Omega \gg \kappa_0, \kappa_1, |\omega - \omega_r|$ increases the frequency mismatch between Floquet replicas, and, thus, the self-energy $\Sigma_{-1}$ becomes a lower-order correction. Specifically, assuming $\mathcal{O}(\Sigma_{-2}) = \Omega^{-1}$, we get

$$\Sigma_{-1} \sim (\cdot)^{-1} \sim \begin{pmatrix} 1/2\Omega & \kappa_0/\Omega^2 \\ \kappa_0/\Omega^2 & 1/\Omega \end{pmatrix} \to 0 \; for \; \Omega \to \infty, \qquad (A13).$$

This behavior is consistent with the observed suppression of transmission to parasitic channels $n < -1$ under fast modulation conditions. In this regime, the effective eigenvalue equation reads

$$\omega \begin{pmatrix} \psi_{n=1,1} \\ \psi_{n=1,2} \\ \psi_{n=0,1} \\ \psi_{n=0,2} \end{pmatrix} = \left[ \begin{pmatrix} \omega_r - \Omega & \kappa_0 & & \kappa_1 \\ \kappa_0 & \omega_r & \kappa_1 & \\ & \kappa_1 & \omega_r & \kappa_0 \\ \kappa_1 & & \kappa_0 & \omega_r + \Omega \end{pmatrix} + \mathcal{O}(\Omega^{-1}) \right] \begin{pmatrix} \psi_{n=1,1} \\ \psi_{n=1,2} \\ \psi_{n=0,1} \\ \psi_{n=0,2} \end{pmatrix} + \begin{pmatrix} 0 \\ 0 \\ i \, 2\gamma_e S_{n=0,1}^+ \\ 0 \end{pmatrix}, \quad (A14).$$

This can be simplified even further by decimating the wave amplitudes $\psi_{n=1,1} = \frac{\kappa_0}{\omega - \omega_r + \Omega} \psi_{n=1,2} + \frac{\kappa_1}{\omega - \omega_r + \Omega} \psi_{n=0,2}$ and $\psi_{n=0,2} = \frac{\kappa_0}{\omega - \omega_r - \Omega} \psi_{n=0,1} + \mathcal{O}(\Omega^{-2})$:

$$\omega \begin{pmatrix} \psi_{n=1,2} \\ \psi_{n=0,1} \end{pmatrix} = \left[ \begin{pmatrix} \omega_r & \kappa_1 \\ \kappa_1 & \omega_r \end{pmatrix} + \mathcal{O}(\Omega^{-1}) \right] \begin{pmatrix} \psi_{n=1,2} \\ \psi_{n=0,1} \end{pmatrix} + \begin{pmatrix} 0 \\ i \, 2\gamma_e S_{n=0,1}^+ \end{pmatrix}, \quad (A15).$$

We now turn to the analysis of the associated scattering problem. In this scenario, the effective scattering matrix for the reduced system is defined in a standard form as

$$\widetilde{S}_F \equiv -\mathbb{I}_2 + 2i\widetilde{D}^T \widetilde{G}_F \widetilde{D}, \qquad (A16)$$

where $\widetilde{D} \equiv \sqrt{2\gamma_e}\mathbb{I}_2$, $\widetilde{G}_F \equiv (\omega \mathbb{I}_2 - \widetilde{H}_{F,\text{eff.}})^{-1}$, and $\widetilde{H}_{F,\text{eff.}} \equiv \begin{pmatrix} \omega_r & \kappa_1 \\ \kappa_1 & \omega_r \end{pmatrix} = \begin{pmatrix} \omega^{(1)} & \kappa_1 \\ \kappa_1 & \omega^{(1)} \end{pmatrix} - i\gamma_e \mathbb{I}_2$. Finally, to realize an SRM, we impose reflectionless scattering boundary conditions at the incident harmonic $n = 0$ from the left lead associated with resonator 1. Specifically, this corresponds to setting the reflection subblock $(\widetilde{S}_F)_{2,2} = 0$. As discussed in Appendix C, this condition collapses the problem of identifying reflectionless modes to an eigenvalue problem involving an auxiliary wave operator. The resulting operator is a non-Hermitian effective Hamiltonian $\widetilde{H}_{SRM} \equiv \begin{pmatrix} \omega^{(1)} + i\gamma_e & \kappa_1 \\ \kappa_1 & \omega^{(1)} - i\gamma_e \end{pmatrix}$, which respects $\mathcal{PT}$-symmetry. This operator corresponds to a $\mathcal{PT}$-symmetric dimer emerging from the decimated Floquet lattice.

## Supplementary Material B: Floquet Analysis for Circuit

This section of the supplementary material details the Floquet analysis as it was applied to the circuit depicted in Fig. 1b and schematically in Fig. 3a. Each transmission line (TL) in the circuit is modeled as an equivalent voltage source with matched impedance $Z_0 = 50\,\Omega$, such that the current $i_m$ and voltage $v_m$ in the TLs can be decomposed into incident and scattered voltages: $v_m = v_m^+ + v_m^-$ and $i_m = \frac{1}{Z_0}(v_m^+ - v_m^-) = \frac{1}{Z_0}(2v_m^+ - v_m)$. Following a similar Fourier series approach as in SM A, assuming a monochromatic input voltage $|v^+\rangle = e^{i\omega t}|V^+\rangle$, where $|v^\pm\rangle \doteq (v_1^\pm \;\; v_2^\pm)^T$, together with periodic modulation of the coupling capacitance $C^\kappa(t) = C^\kappa(t + 2\pi/\Omega) = \sum_l C_l^\kappa e^{il\Omega t}$, we model the system response as $|\varphi(t)\rangle = \sum_n e^{i(\omega+n\Omega)t}|\Phi_n\rangle$, where the components of $|\varphi\rangle \doteq (u_1 \;\; u_2 \;\; v_1 \;\; v_2)^T$ are the voltages of the circuit and $|\Phi_n\rangle \doteq (U_{1,n} \;\; U_{2,n} \;\; V_{1,n} \;\; V_{2,n})^T$. We also have $j_{21}(t) = \sum_n e^{i(\omega+n\Omega)t} J_{21,n}$, which is omitted from $|\varphi\rangle$ as it will be eliminated momentarily. Using Kirchhoff's voltage and current laws, the equations associated with the time-independent elements are readily obtained and summarized in the frequency domain, after eliminating branch currents in each resonator and in the TLs,

$$\alpha_m V_{m,0} - U_{m,0} = 2\frac{Z_m^e}{Z_0} V_{m,0}^+ \quad (B1a)$$

$$(-1)^{m-1} Z_0 J_{21,0} + \frac{Z_0}{Z_m} U_{m,0} + V_{m,0} = 2V_{m,0}^+ \quad (B1b)$$

for $m = 1,2$, where $Z_m(\omega) = i\frac{\omega}{C_m}\frac{1}{\omega_{0,m}^2 - \omega^2}$ is the impedance of each LC resonator with their isolated resonant angular frequencies $\omega_{0,m} = \frac{1}{\sqrt{LC_m}}$. The impedance of the capacitors connecting the TLs is $Z_m^e(\omega) = \frac{1}{i\omega C_m^e}$, and $\alpha_m(\omega) \equiv 1 + \frac{Z_m^e(\omega)}{Z_0}$. The equation associated with the time-dependent capacitor reads

$$j_{21}(t) = \frac{d}{dt}\left[C^\kappa(t)\bigl(u_1(t) - u_2(t)\bigr)\right] \quad (B2)$$

Inserting the Fourier series for each term and employing the same orthogonality trick used in the CMT formulation to balance the harmonics, after simplification the result is

$$J_{21,0}(\omega) = \sum_l \frac{1}{Z_l^\kappa(\omega)}\left[U_{1,-l} - U_{2,-l}\right] \quad (B3)$$

where $Z_l^\kappa(\omega) = \frac{1}{i\omega C_l^\kappa}$. Eliminating $J_{21,0}$ by combining this with the previous frequency domain Eqs. $(B1)$, the equations for harmonic $n = 0$ can be concisely expressed,

$$\sum_l M_F^{(l)}(\omega)|\Phi_{-l}\rangle = A(\omega)|V_0^+\rangle \quad (B4)$$

with the sum over $l$ describing the interaction between harmonics, and where $|V_0^+\rangle \doteq (V_{1,0}^+, V_{2,0}^+)^T$. Generally, for the $n^{th}$ harmonic, in which the arguments of Eq. $(B4)$ are evaluated at the associated frequency,

$$\sum_l M_F^{(l)}(\omega + n\Omega)|\Phi_{n-l}\rangle = \delta_{n0} A(\omega + n\Omega)|V_0^+\rangle \quad (B5)$$

where the matrices can be represented in the form

$$A(\omega) \doteq 2\begin{pmatrix} Z_1^e/Z_0 & 0 \\ 0 & Z_2^e/Z_0 \\ 1 & 0 \\ 0 & 1 \end{pmatrix} \quad (B6a)$$

$$M_F^{(0)}(\omega) \doteq \begin{pmatrix} -1 & 0 & \alpha_1 & 0 \\ 0 & -1 & 0 & \alpha_2 \\ \beta_1 & -\rho_0 & 1 & 0 \\ -\rho_0 & \beta_2 & 0 & 1 \end{pmatrix} \quad (B6b)$$

$$M_F^{(l \neq 0)}(\omega) \doteq \rho_l \begin{pmatrix} 0 & 0 & 0 & 0 \\ 0 & 0 & 0 & 0 \\ 1 & -1 & 0 & 0 \\ -1 & 1 & 0 & 0 \end{pmatrix} \quad (B6c)$$

with $\rho_l \equiv Z_0/Z_l^\kappa$ and $\beta_m \equiv \rho_0 + Z_0/Z_m$.

As in SM A, we can generalize the formulation to input waves at arbitrary harmonics by introducing the extended-space notation, where now $f$ denotes either $\Phi$ or $V^\pm$. This way, we can express the system in the matrix form

$$M_F \vec{\Phi} = A_F \vec{V}^+ \quad (B7)$$

where $M_F$ and $A_F$ have the harmonic-indexed blocks,

$$[M_F(\omega)]_{p,q} = M_F^{(q-p)}(\omega - p\Omega) \quad (B8a)$$
$$[A_F(\omega)]_{p,q} = \delta_{pq} A(\omega - p\Omega) \quad (B8b)$$

By introducing the matrix $W \doteq \begin{pmatrix} 0 & 0 & 1 & 0 \\ 0 & 0 & 0 & 1 \end{pmatrix}$, defined such that $|V\rangle = W|\Phi\rangle$, and reusing the tilde notation of SM A, we can express the input/output relation for arbitrary harmonics,

$$\vec{V}^- = -\vec{V}^+ + \widetilde{W}\vec{\Phi} \quad (B9)$$

Hence, Eqs. (B8) and (B9) can be used to express the Floquet scattering matrix, which connects the input/output voltages between arbitrary harmonics $\vec{V}^- = S_F \vec{V}^+$,

$$S_F = -\widetilde{\mathbb{I}}_M + \widetilde{W} M_F^{-1} A_F \quad (B10)$$

where $\mathbb{I}_M$ is an identity matrix with the same dimensionality as $|V^\pm\rangle$. As was the case in SM A, $M_F$ must be truncated to a finite number of harmonics to be numerically inverted.

### Supplementary Material C: Auxiliary Floquet Lattice

The previous sections of the supplementary material provided frequency-domain Floquet analysis for CMT and circuit formulations, respectively. In this section, we use those formulations to demonstrate how specified Floquet scattering boundary conditions—e.g., reflectionless frequency conversion—can be imposed to define an auxiliary wave operator that naturally encodes the desired solutions. We then provide a physical interpretation of the system described by the auxiliary operator.

Both formulations were cast in matrix form in Eqs. (A2) and (B7), alongside their input/output relations in Eqs. (A4) and (B9), respectively. Here, we follow the notation of SM B, but the methodology is general. The auxiliary wave operator is obtained naturally when the frequency-domain equations are written in the form,

$$M_{SRZ} \vec{\Phi} = A_{SRZ} \vec{V}_{SRZ} \quad (C1)$$

where the scattering boundary conditions are enforced by setting $\vec{V}_{SRZ} = 0$. Therefore, non-trivial solutions $\vec{\Phi}$ exist when $\det M_{SRZ}(\omega_{SRZ}) = 0$. Floquet reflectionless scattering occurs when the system is driven by a source with central harmonic at $\omega = \omega_{SRZ}$, and appropriate incident wavefront—potentially multiplexing different input harmonics. This wavefront is determined by the null vectors of $M_{SRZ}$, together with the input/output relation, Eq. (B9), which is trivial in the case of a single input channel. As always, steady-state solutions require the driving frequency to be real, i.e., $\omega_{FRZ} \in \mathbb{R}$. In the case of reflectionless boundary conditions, these solutions are termed synthetic reflectionless scattering modes (SRMs), while solutions with $\omega_{FRZ} \in \mathbb{C}$ are generally referred to as synthetic reflection zeros.

#### *1. Synthetic Reflection Zeros in Circuit*

For analysis of the synthetic reflection zeros in the circuit discussed in the main text, reflectionless boundary conditions for a wave incident at the first TL in harmonic $n = 0$ are encoded in the vector $\vec{V}_{FRZ} =$

$(\ldots \ |V_1^+\rangle \ |V_0^{RZ}\rangle \ |V_{-1}^+\rangle \ \ldots)^T = 0$, where the $n=0$ components $|V_0^{RZ}\rangle \doteq (V_{1,0}^- \ V_{2,0}^+)^T$ consist of the reflected voltage amplitude at the input TL and incident voltage at the second TL. Only the two equations in $(B1)$ corresponding to the input TL require modification. Given that $V_{1,0}^+ = -V_{1,0}^- + V_{1,0}$, we can rewrite them as

$$\tilde{\alpha}_1 V_{1,0} - U_{1,0} = -2\frac{Z_1^e}{Z_0} V_{1,0}^- \quad (C2a)$$

$$-Z_0 J_{21} - \frac{Z_0}{Z_1} U_{1,0} + V_{1,0} = 2V_{1,0}^- \quad (C2b)$$

where $\tilde{\alpha}_1(\omega) \equiv 1 - \frac{Z_1^e(\omega)}{Z_0}$. Accordingly, the modified blocks at the harmonic $n=0$ are,

$$A_{SRZ}^{(0)}(\omega) \doteq 2\begin{pmatrix} -Z_1^e/Z_0 & 0 \\ 0 & Z_2^e/Z_0 \\ 1 & 0 \\ 0 & 1 \end{pmatrix} \quad (C3a)$$

$$M_{SRZ}^{(0)}(\omega) \doteq \begin{pmatrix} -1 & 0 & \tilde{\alpha}_1 & 0 \\ 0 & -1 & 0 & \alpha_2 \\ -\beta_1 & \rho_0 & 1 & 0 \\ -\rho_0 & \beta_2 & 0 & 1 \end{pmatrix} \quad (C3b)$$

the matrix subblocks in $(C1)$ can be written,

$$[A_{SRZ}(\omega)]_{p,q} = \begin{cases} A_{SRZ}^{(0)}, & p=0 \\ [A_F]_{p,q}, & p \neq 0 \end{cases} \quad (C4a)$$

$$[M_{SRZ}(\omega)]_{p,q} = \begin{cases} M_{SRZ}^{(0)}, & p=0 \\ [M_F]_{p,q}, & p \neq 0 \end{cases} \quad (C4b)$$

The system described by $(C1)$ can be interpreted physically in relation to the system described by $(B7)$. The only difference is a sign change in the characteristic impedance of the TL at the input harmonic.

### *2. Reflection Zeros in CMT*

Elimination of the input variables in favor of the corresponding output variables in the framework of CMT results in an auxiliary Floquet Hamiltonian, which—similar to the previous circuit example—can be interpreted in relation to the effective Hamiltonian of the synthetic lattice. Consider the frequency-domain equations at an input TL and harmonic $\alpha$ with external coupling strength $\gamma_e$,

$$\omega \psi_\alpha = (\varepsilon_\alpha - i\gamma_e)\psi_\alpha + i\sqrt{2\gamma_e} s_\alpha^+ + \cdots \quad (C5)$$

where the ellipses refer to additional coupling terms not explicitly containing $\psi_\alpha$ or $s_\alpha^+$. Using the input/output relation from Eq. $(A4)$,

$$s_\alpha^+ = -s_\alpha^- + \sqrt{2\gamma_e}\psi_\alpha \quad (C6)$$

We can rewrite Eq. $(C5)$ as

$$\omega \psi_\alpha = (\varepsilon_\alpha + i\gamma_e)\psi_\alpha - i\sqrt{2\gamma_e} s_\alpha^- + \cdots \quad (C7)$$

where $s_\alpha^- = 0$ is enforced by the reflectionless boundary conditions. This equation is part of the eigenvalue problem defined in Eq. (4) of the main text, whose eigenstates describe reflectionless solutions. From Eq. $(C7)$, the auxiliary Floquet Hamiltonian is obtained by reversing the sign of the coupling to the input TL/harmonic in the effective Hamiltonian of the synthetic lattice. In the effective Hamiltonian, the external coupling was interpreted as an equivalent loss. Thus, by reversing its sign, the effect can be interpreted as an equivalent gain in the auxiliary system. This explains why the second panel of Fig. 1a depicts a red site at the input TL and harmonic, while the others are green.

In the case of CMT modeling, the resulting auxiliary system is naturally cast as an eigenvalue problem, whereas in conventional scenarios, e.g., electronic circuits, the reflectionless solutions are obtained by identifying real-frequency zeros of an auxiliary wave operator as in $(C1)$. As a final note, we point out that solutions satisfying the Floquet reflectionless scattering boundary conditions do not necessarily guarantee perfect transmission to the chosen channels/harmonics. Rather, in the cases presented in this work, efficient and broadened frequency conversion was enabled by the tailored connectivity of the Floquet ladder.

### 3. Degeneracy of Reflection Zeros

The substitutions performed in Eqs. ($C2$) and ($C7$) are particular cases of a generic methodology, which can be formalized by introducing filtering matrices to partition the incident and scattered wave variables into disjoint subspaces depending on the chosen Floquet scattering boundary conditions. Specifically, denoting the variables of the input and output channels with subscripts '$in$' and '$out$', respectively, again following the notation of SM B, we write the decomposition,
$$\vec{V}^{\pm} = F_{in}^T \vec{V}_{in}^{\pm} + F_{out}^T \vec{V}_{out}^{\pm} \quad (C8)$$
where the chosen boundary conditions are satisfied when $\vec{V}_{out}^+ = \vec{V}_{in}^- \equiv 0$. The filtering matrices in Eq. ($C8$) satisfy $F_{in}^T F_{in} + F_{out}^T F_{out} = \tilde{\mathbb{I}}_M$, $F_{in} F_{in}^T = \mathbb{I}_{in}$, $F_{out} F_{out}^T = \mathbb{I}_{out}$, and $F_{out} F_{in}^T \equiv 0$, such that they project $\vec{V}^{\pm}$ onto their respective subspaces $\vec{V}_{in}^{\pm} = F_{in} \vec{V}^{\pm}$ and $\vec{V}_{out}^{\pm} = F_{out} \vec{V}^{\pm}$. The identity matrices, $\mathbb{I}_{in}$ and $\mathbb{I}_{out}$ have the same dimensionality as $\vec{V}_{in}^{\pm}$ and $\vec{V}_{out}^{\pm}$, respectively. Together with the input/output relation ($B9$), the filtering matrices can be used to rewrite the system equations ($B7$) in the form
$$(M_F - A_F F_{in}^T F_{in} W) \vec{\Phi} = -A_F (F_{in}^T \vec{V}_{in}^- - F_{out}^T \vec{V}_{out}^+) \quad (C9)$$
which is equivalent to ($C1$) with,
$$M_{SRZ} \equiv B(M_F - A_F F_{in}^T F_{in} W) \quad (C10)$$
where $B$ is a diagonal matrix with diagonal entries $\pm 1$ depending on the choice of representation for $M_{FRZ}$.

From the Floquet scattering matrix of Eq. ($B10$), filtering matrices naturally extract the Floquet reflection matrix,
$$\hat{r}_F = -\mathbb{I}_{in} + F_{in} \widetilde{W} M_F^{-1} A_F F_{in}^T \quad (C11)$$
Computing the determinant and applying the matrix determinant lemma—$\det T + UV^T = \det T \det \tilde{\mathbb{I}}_N + V^T T^{-1} U$—with $T = -\mathbb{I}_{in}$, $U = F_{in} \widetilde{W}$ and $V^T = M_F^{-1} A_F F_{in}^T$,
$$\det \hat{r}_F = \det -\mathbb{I}_{in} \det \tilde{\mathbb{I}}_N - M_F^{-1} A_F F_{in}^T F_{in} \widetilde{W} \quad (C12)$$
Factoring out $M_F^{-1}$ and recognizing the form of ($C11$),
$$\det \hat{r}_F = \pm \frac{\det M_{SRZ}}{\det M_F} \quad (C13)$$
where the sign depends on the representation choice through $B$ and the dimensionality of $\hat{r}_F$.

In the case of only a single input channel, the reflection matrix reduces to a reflection coefficient. By also assuming $M_{FRZ}$ possesses a second-order zero at real frequency $\omega_0$ that is separated from resonance poles of the system (zeros of $M_F$), the reflection coefficient scales as
$$r_F \sim (\omega - \omega_0)^2 \quad (C14)$$
which leads to quartic scaling of reflectance with respect to frequency detuning $\nu \equiv \omega - \omega_0$,
$$R \sim \nu^4 \quad (C15)$$
This reasoning also extends to the CMT formulation.

### Supplementary Material D: Circuit Parameters

This section of the supplementary material reports approximate parameter values used in each simulation configuration. The truncated precision introduces only marginal qualitative differences to the results of the main text. Both configurations were simulated including seven harmonics to ensure convergence. Parameters that are common to both configurations are as follows: $Z_0 = 50 \, \Omega$, $L = 220 \, \mu H$, $C_1 = 1638$ pF, $C_1^e \approx 739$ pF and $C_0^\kappa = 96$ pF. Parameters unique to the results of Fig. 3c are $\Omega \approx 2\pi \times 76$ kHz, $C_2 \approx 730$ pF, $C_2^e \approx 538$ pF and $\eta \approx 0.36$. Parameters unique to the results of Fig. 3d are $\Omega \approx 2\pi \times 23$ kHz, $C_2 \approx 1303$ pF, $C_2^e \approx 660$ pF and $\eta \approx 0.38$. Prior to their iterative refinements, circuit parameters were initialized to abide by the symmetry constraints informed by our CMT modeling. Specifically, following the notation of SM B,
$$Z_1(\omega_0) = Z_2(\omega_0 + \Omega) \quad (D1a)$$

$$Z_1^e(\omega_0) = Z_2^e(\omega_0 + \Omega) \quad (D2a)$$

where $\omega_0 \approx 2\pi \times 215$ kHz is an estimate of the central operating frequency of the incident wave in proximity to the second-order reflection zero.